\documentstyle[aps,prl,graphicx,floats]{revtex}

\begin{document}

\draft
\title{Helix Formation and Folding in an Artificial Peptide}   
\author{ Nelson A. Alves\footnote{E-mail: alves@quark.ffclrp.usp.br}}
\address{Departamento de F\'{\i}sica e Matem\'atica, FFCLRP
     Universidade de S\~ao Paulo. Av. Bandeirantes 3900. 
     CEP 014040-901 \, Ribeir\~ao Preto, SP, Brazil}
\author{Ulrich H.E. Hansmann \footnote{hansmann@mtu.edu, 
                        to whom all correspondence should be addressed}}
\address{Department of Physics, Michigan Technological University,
         Houghton, MI 49931-1291, USA}
\date{\today}
\maketitle
\begin{abstract}
We study the relation between $\alpha$-helix formation and 
folding  for a simple artificial peptide, Ala$_{10}$-Gly$_5$-Ala$_{10}$. 
Our data rely on multicanonical Monte Carlo simulations 
where the interactions among all atoms are taken into account. 
The free-energy landscape of the peptide is evaluated for various 
temperatures. Our data indicate that folding of this peptide is a
two-step process: in a first step two $\alpha$-helices are formed which
afterwards re-arrange themselves  into a U-like structure.
\end{abstract}

\section{Introduction}
The mechanism by which a large class of proteins folds spontaneously
into  a unique  globular shape \cite{Anf} has remained elusive. 
Significant new insight was gained over the last few years from the 
studies of minimal protein models. For instance,  energy landscape theory
and funnel concept \cite{Bryngelson87,Onuchic97} proved to be
powerful tools for description of the general characteristics  of folding
not only in  minimalistic protein models  but also for real proteins 
\cite{HOO98b,HO01b}. However, many questions
on the details of the folding process remain to be solved. For instance,
folding of proteins involves one or more transitions  between  different 
thermodynamic states. The role of these transitions in the folding
process is an active area of research. An important example for
these transitions is  the formation of secondary 
structure elements.
 For the case of $\alpha$-helices it is long known that
there is a sharp transition towards a random coil state when the 
temperature is increased. The characteristics of this so-called 
helix-coil transition have been studied extensively \cite{Poland},
most recently in Refs.~\cite{Jeff,HO98c,AH99b,AH00b,MO2,PH01f}.
In this paper, we  research  the relation between helix-coil transition 
and folding. 

For this purpose, we have studied an artificial peptide, 
Ala$_{10}$-Gly$_5$-Ala$_{10}$, in a detailed representation 
where the interactions between all atoms are taken into account. 
Multicanonical simulations \cite{MU} with large statistics are 
used to evaluate the free energy landscape of our peptide at 
different temperatures.  The encountered transitions are further
investigated 
by partition zeros analysis which allows to characterize 
``phase transitions'' in small systems \cite{BMH}.  Quantities such as 
the energy, specific heat, helicity and susceptibility were calculated 
as function of temperature. 
We have neglected in the simulations the interaction of our artifical
peptide with the surrounding solvent. While this is certainly
a crude approximation, it allows us not only to relate our results 
to our previous studies  on helix-coil transition
in poly-alanine that also relied on gas-phase simulations 
\cite{HO98c,AH99b,AH00b}, but also  
 to study the extend to that secondary structure formation and
folding are determined by intrinsic  properties of the peptide.
Our data suggest that the peptide  in gas-phase folds 
in a two-step process: in a first step two $\alpha$-helices are formed 
in what amounts to a first order transition. Afterwards these helices 
re-arrange themselves  into a U-like structure. 
The second step has the characteristics of a second order transition.


\section{Methods}
Our investigation of  Ala$_{10}$-Gly$_5$-Ala$_{10}$ is based on a detailed, 
all-atom representation of that peptide. The interaction between the 
atoms is described by a standard force field, ECEPP/2,\cite{EC}  
(as implemented in the  program package SMMP \cite{SMMP}) and is given by:
\begin{eqnarray}
E_{tot} & = & E_{C} + E_{LJ} + E_{HB} + E_{tor},\\
E_{C}  & = & \sum_{(i,j)} \frac{332q_i q_j}{\epsilon r_{ij}},\\
E_{LJ} & = & \sum_{(i,j)} \left( \frac{A_{ij}}{r^{12}_{ij}}
                                - \frac{B_{ij}}{r^6_{ij}} \right),\\
E_{HB}  & = & \sum_{(i,j)} \left( \frac{C_{ij}}{r^{12}_{ij}}
                                - \frac{D_{ij}}{r^{10}_{ij}} \right),\\
E_{tor}& = & \sum_l U_l \left( 1 \pm \cos (n_l \chi_l ) \right).
\end{eqnarray}
Here, $r_{ij}$ (in \AA) is the distance between the atoms $i$ and $j$,
 and $\chi_l$ is the $l$-th torsion angle. The peptide bond angles 
were set to their common value $\omega = 180^{\circ}$. We do not  
include explicitly  the interaction of the peptide with the solvent 
into our simulations and set the dielectric constant $\epsilon$ equal to 2.
Since the charges at peptide termini are known to
reduce helix content \cite{SKY}, we removed them by taking a 
neutral NH$_2$-- group at the
N-terminus and a neutral --COOH group at the C-terminus.

Simulation of detailed protein models where the interaction between all
atoms are taken into account are extremely difficult. This is because 
the various competing interactions within the molecule lead to an energy 
landscape characterized by a multitude of local minima separated by
high energy barriers.  Hence, in the low-temperature region,
canonical Monte Carlo or molecular dynamics simulations will tend to get 
trapped in one of these minima and the simulation will not thermalize 
within the available CPU time. One example of the new and sophisticated
algorithms \cite{curr_op} that allow to overcome this difficulty are
{\it generalized-ensemble} techniques \cite{HO,OurReview98}, and 
it is one of these techniques, multicanonical 
sampling \cite{MU}, that we used for our investigations. 

In the multicanonical algorithm \cite{MU}
conformations with energy $E$ are assigned a weight
$  w_{mu} (E)\propto 1/n(E)$. Here, $n(E)$ is the density of states.
A  simulation with this weight
will  lead to a uniform distribution of energy:
\begin{equation}
  P_{mu}(E) \,  \propto \,  n(E)~w_{mu}(E) = {\rm const}~.
\label{eqmu}
\end{equation}
This is because the simulation generates a 1D random walk in the
energy space,
allowing itself to escape from any  local minimum.
Since a large range of energies are sampled, one can
use the reweighting techniques \cite{FS} to  calculate thermodynamic
quantities over a wide range of temperatures $T$ by
\begin{equation}
<{\cal{A}}>_T ~=~ \frac{{\int dx~{\cal{A}}(x)~w^{-1}(E(x))~
                 e^{-\beta E(x)}}}
              {{\int dx~w^{-1}(E(x))~e^{-\beta E(x)}}}~,
\label{eqrw}
\end{equation}
where $x$ stands for configurations.

Note that unlike in the case of canonical simulations the weights
\begin{equation}
w(E) = n^{-1}(E) = e^{-S(E)}
\label{eqweight}
\end{equation}
are not a priori known. Instead estimators for these weights have to
be determined  by  an iterative procedure \cite{Berg,PH01f}. In our
case we needed 500,000  sweeps for the  weight factor calculations. 
All thermodynamic quantities were  then estimated from one production 
run of $8,000,000$ Monte Carlo sweeps which followed 10,000 sweeps for
thermalization.  Our simulations were started from  completely random 
initial conformations (Hot Start) and  one Monte Carlo sweep updates 
every torsion angle of the peptide once.  At the end of every 10th 
sweep we stored the ECEPP/2 energies $E_{tot},E_C, E_{LJ}, E_{hb}$ 
and $E_{tor}$ of the conformation,  the corresponding number $n_H$ 
of helical residues and end-to-end distance $d_{e-e}$. 
Here, we follow  previous work \cite{OH95b} 
and consider a residue as helical if its backbone angle $(\phi,\psi)$ are 
within the range $(-70^{\circ}\pm 20^{\circ},-37^{\circ}\pm20^{\circ})$.

Using the results of our generalized-ensemble simulation, we explored for
various temperatures the free energies
\begin{equation}
G(n_H,d_{e-e}) = -k_B T \log P(n_H,d_{e-e})~.
\end{equation}
Here, $P(n_H,d_{e-e})$ is the probability to find a peptide conformation 
with values $\ell$, $d_{e-e}$ (at temperature $T$). We chose the 
normalization so that the lowest value of $G(n_H,d_{e-e})$ is set to 
zero for each temperature.

We finally used that the multicanonical algorithm allows us
to calculate estimates for the spectral density:
\begin{equation}
  n(E) = P_{mu} (E) w^{-1}_{mu} (E)~.
\end{equation}
We can therefore construct
the corresponding  partition function for our all-atom model of
 Ala$_{10}$-Gly$_5$-Ala$_{10}$ from these estimates by
\begin{equation}
     Z(\beta) = \sum_{E} n(E) e^{- \beta E} ,                   \label{eq:r1}
\end{equation}
with $\beta$ the inverse temperature, $\beta = 1/k_B T$.
The complex solutions of the partition function
determine the critical behavior of the model and were also studied by us.


\section{Results and Discussion}
Our peptide,  Ala$_{10}$-Gly$_5$-Ala$_{10}$, is build up out of two
chains of each 10 alanine residues connected by 5 glycine residues. 
In previous work \cite{OH95b,HO98c,AH99b} we could show that 
polyalanine has a pronounced transition between a disordered 
coil phase and an ordered state in which the polymer forms 
an $\alpha$-helix. For this reason, we  expect  formation 
of $\alpha$-helices in our peptide, and the average number of helical 
residues $<n_H>$ is therefore one of the quantities
that we have measured.  $<n_H>$ is displayed in Fig.~1 as a function of 
temperature, and we observe in this plot  two temperature
regions. At high temperature, few  residues are found with backbone
dihedral angles $(\phi,\psi)$ typical for an $\alpha$-helix. On the 
other hand, at low temperatures we observe helix-formation,
and almost all of the alanine residues are part of an $\alpha$-helix, 
i.e. have backbone dihedral angles  $(\phi,\psi)$ in the range 
($-70^{\circ}\pm 20^{\circ},-37^{\circ}\pm 20^{\circ}$). 
The transition between the two temperature
regions is sharp indicating the existence of a helix-coil transition.
The  transition temperature $T_{hc}$ can be determined from the
corresponding peak in the susceptibility
\begin{equation}
 \chi(T) = <n^2_H(T)> - <n_H(T)>^2 \, ,
\end{equation}
which is ploted in the inset of Fig.~1, and we find the
transition temperature $T_{hc} = 485\pm 5$ K. 

In previous work \cite{OH95b,PH01f} we could show that in polyalanine 
the formation of $\alpha$-helices is related to a gain in potential 
energy. For this reason, we display in Fig.~2  the average total 
ECEPP/2 energy $<E_{tot}>$ and the thermodynamic averages of partial 
energies $<E_C>$, $<E_{LJ}>$, $<E_{hb}>$ and $<E_{tor}>$ of
Ala$_{10}$-Gly$_5$-Ala$_{10}$ as a function of temperature.
As expected, we observe around our transition temperature $T_{hc}$
a sharp decrease in the potential energy $<E_{tot}>$ that  is  due to 
a corresponding decrease   in the Coulomb energy $<E_C>$, 
Lennard-Jones energy $<E_{LJ}>$ and hydrogen-bond energy $<E_{hb}>$.
The change in  $<E_{tot}>$ with temperature can be described by the 
specific heat
\begin{equation}
  C(T)  = \beta^2 \ \frac{<E_{tot}^2> - <E_{tot}>^2}{25},
\end{equation}
which we display in Fig.~3.  A pronounced change in energy with 
temperature corresponds to a peak in the specific heat.  As one 
can see from Fig.~3, we observe indeed a pronounced peak in the 
specific heat at a temperature $T=480 \pm 10$ K that is consistent
with $T=485 \pm 5$ K,  the temperature where the peak in the 
susceptibility is located.  Combining both values we obtain as our 
final estimate for the  helix-coil transition temperature 
$T_{hc}=483\pm 8$. However, we find in Fig.~3 also a second, smaller 
peak in the specific heat at the lower temperature $T_f=265\pm 7$ K 
indicating yet another transition.

The sharp decrease in potential energy corresponding to
this second peak is clearly  visible in Fig.~2 for the total energy
 $<E_{tot}>$, however, not for all of the partial energies. Only the
Lennard-Jones term $<E_{LJ}>$ exhibits also a signal for the second
transition at $T_f$.  This energy term depends strongly on the
overall size of the molecule and the change in this quantity
indicates a transition between extended and compact structures.
Hence, we conjecture that the second peak in specific heat at the
lower temperature $T_f$ is related to a transition between extended
and compact structures.  A possible measure for such a change 
is the average end-to-end distance $<d_{e-e}>_T$.  We define here 
 $<d_{e-e}>$ as the distance between N of Ala$_1$ and O of Ala$_{25}$,
and plot this quantity in Fig.~4. We observe that this quantity decreases
with decreasing temperature. Below the helix-coil transition $T_{hc}$
the decrease slows down and the curve becomes almost flat at a value
of $<d_{e-e}> \approx 10$ \AA \ indicating that
there is little further change in the compactness of the molecule. 
However, at temperature $T_f$  the end-to-end distance decreases
again sharply towards a new value $<d_{e-e}> = 6.1$ \AA .  
Hence, $T_f$ marks the folding of the molecule into  a defined 
compact structure  with 
the two terminal ends of the peptide  close together.

The transition between extended and a more compact structure can also 
be seen when we display the free energy landscape of our peptide
as a function of helicity $n_H$ and end-to-end distance $d_{e-e}$.
At the temperature $T=480$ K (which is essentially the helix-coil 
transition temperature $T_{hc}= 483\pm 8$ K) the free energy 
landscape (displayed in Fig.~5a) is flat over a large range of values
of $n_H$ and $d_{e-e}$. The $3 k_BT$ contour line surrounds a region where
the helicity can take values between $0 \le n_H \lesssim  20$ and
the end-to-end distance values between $3 \lesssim d_{e-e} \lesssim 40$,
allowing the system to move freely between extended and compact
configurations, and between helical and coil configurations.
On the other hand, at the second and lower  temperature 
 $T=270$ K (which is essentially the  folding transition
temperature $T_f=265\pm 7$ K) the free energy grows rapidly with 
decreasing helicity $n_H$ favoring configuration in a small strip 
with $ 15 \lesssim n_H \lesssim 25$. Hence, the plot of the free
energy landscape in Fig.~5b is limited to  values $12 \lesssim 
n_H \lesssim 25$ of the helicity. Here,  two regions of minimal 
free energy can be seen (marked by the 3 $k_BT$ contour lines). 
The first minima is found at values of $d_{e-e}$ between 5 and 10 \AA \  
and $15 \le n_H \le 20$   characterising compact structures. 
A second region with slightly lower free energy (see the 1 $k_BT$ 
contour line) is found  at much larger values of $d_{e-e}$ 
between 35 and 40 \AA \ and $ 20 < n_H \le 25$ indicating a long 
stretched $\alpha$-helix. Both local free energy minima are 
separated by free energy barriers of  $\approx 8$ $k_BT$  that 
can be  overcome by thermal fluctuations. On the other hand,
configurations with helicity $n_H < 10$ are suppressed by free 
energy differences of more than 30 $k_BT$.

Examples for the structures corresponding to the two free energy minima
are plotted in Fig.~6. The first one, displayed in Fig.~6a is the 
configuration with lowest energy ever found in our multicanonical
simulation of 8,000,000 sweeps and corresponds to the region in
the free-energy landscape at  values of $d_{e-e}$ between 5 and 10 \AA \
and $15 \le n_H \le 20$. This conformation (`A') consists  out
of two helixes (made up out of the alanine residues) connected by
a turn (build out of the flexible glycine residues) towards a 
U-turn-like structure that is consistent with the small value of
the end-to-end distance $d_{e-e}$ observed in Fig.~4 for 
temperatures below $T_f$.  For reference we show  in Fig.~6b 
also the configuration (`B') where all
25 residues are part of an $\alpha$-helix and which corresponds to
the second local free-energy minimum in Fig.~5b at values of
 $d_{e-e}$ between 35 and 40 \AA \ and $ 20 < n_H \le 25$ . The  
dihedral angles of both configurations are listed in Table 1.  
Fig.~7 displays the frequency of both 
configurations as a function of temperature.  For $T > T_{hc}$ neither
configuration `A' nor `B' are observed. Below that temperature 
both structures appear with similar probability as long as the temperature
is higher than $T_f$. At $T=T_f$ the probability to find the maximal
helical structure `B' is with $\approx 30$ \% highest and decreases after that
with decreasing temperature. On the other hand, the frequency
for the U-turn structure `A' continues to grow with decreasing temperature. 
This different behavior is due to the energy differences between
both structures. The minimal
energy conformation `A' has with $E_{tot} =-34.7$ Kcal/mol a $10.8 $ Kcal/mol
lower potential energy than the extended helix conformation `B'
($E_{tot}=-23.9 $ Kcal/mol). This difference is mainly due to the
Lennard-Jones terms: $E_{LJ}=-132.5$ Kcal/mol for `A' 
vs. $E_{LJ}=-118.9$ Kcal/mol
for `B'. The gain in $E_{LJ}$ is  in part compensated by the
hydrogen-bonding terms: $E_{hb}=-30.2$ Kcal/mol  for `A' vs. $E_{hb}=-34.7$
Kcal/mol for `B'. Coulomb and torsion energies differed little between
the two configurations: $E_C=126.7$ Kcal/mol and $E_{tor}=1.3$ Kcal/mol for
conformation `A' vs. $E_C=126.4$ Kcal/mol and $E_{tor}=3.3$ Kcal/mol for
conformation `B'. 

It is an interesting question whether our two observed transitions
(occuring in a finite and small system) can be
related to phase transitions which in a strict sense are defined only
for macroscopic (that is very large) systems. In order to study this
question we have  calculated the 
complex zeros $\beta \rightarrow {\rm Re}(\beta) + i \tau$
of the partition function $Z(\beta)$ of our molecule.  
In the case of a temperature driven phase transition, 
we expect that the complex zeros
$\beta_j,~(j=1,2, ...)$
(or at least the ones close to the real axis) 
condense for large enough system size  on a single line.
As the system size  increases, those zeros will move towards
the positive real $\beta$-axis  and the corresponding value is for large
system size  the inverse of the physical critical temperature $T_c$.
Crucial information on phase transitions can be obtained from
the way in which the first zero approaches the real $\beta$-axis.
However, such an analysis depends on the extrapolation
towards the infinite large system and does not allow  characterization
of the situation in small systems such as Ala$_{10}$-Gly$_5$-Ala$_{10}$.
One possible extension of the above ideas to ``phase transitions'' in
biological molecules and other small systems is the classification 
scheme by  Borrmann {\it et al.} \cite{BMH}. 
 In this approach one computes the discrete line density of zeros as
an average of the inverse distances between neighboring zeros,
\begin{equation}
\phi(\tau_k) = \frac{1}{2} \left(\frac{1}{|\beta_k - \beta_{k-1}|}
       + \frac{1}{|\beta_{k+1} - \beta_{k}|}\right)\,,     \label{eq:phi}
\end{equation}
and  approximates $\phi(\tau)$ by a simple power law
$\phi(\tau) \sim \tau^{\alpha}$. Taking the first four complex zeros,
one obtains
\begin{equation}
 \alpha = \frac{ {\rm ln}\,\phi(\tau_3) - {\rm ln}\,\phi(\tau_2)}
               { {\rm ln}\,\tau_3 - {\rm ln}\,\tau_2} \, .    \label{alpha}
\end{equation}
With a second parameter $\gamma$, related to the crossing angle of this
line with the real axis,  
\begin{equation}
  \gamma = 
  [{\rm Re}(\beta_2) -{\rm Re}(\beta_1)]/(\tau_2 - \tau_1)\,, \label{gamma}
\end{equation}
and following the classification scheme by Grossmann
and Rosenhauer \cite{GR1,GR2},
phase transitions can now be classified according to the values of these
two parameters:
for $\alpha \leq 0$ and $\gamma=0$ one has a phase transition of
first order, it is of second order if $0 < \alpha < 1$ and
arbitrary $\gamma$, and for $\alpha > 1$
and arbitrary $\gamma$ one has a higher order transition. We have
evaluated the usefulness of this approach both for spin systems and
polyalanine chains \cite{AFH01d,AHP01e}. Preliminary results for
 Ala$_{10}$-Gly$_5$-Ala$_{10}$ are also listed in Ref.~\cite{AHP01e}.

For Ala$_{10}$-Gly$_5$-Ala$_{10}$, we  find  two lines of complex zeros.
The corresponding first four zeros for each characteristic line are 
listed in Table 2. Our error estimate is based on the jackknife 
method \cite{jack} with 16 bins. These lines lead to  two critical 
temperatures $T_{hc}=480$ K  and $T_f=271$ K (estimated from the real
part of $\beta$) that agree  with 
the corresponding values $T_{hc}=483 \pm 8$ K and $T_f=265 \pm 7$ K
found  by us above with different methods.

Using these zeros  we  have calculated
the parameters $\alpha$ and $\gamma$ that characterize
in the Borrmann {\it et al.} approach phase transitions in small systems.
For the first transition, at $T=480$ K,
we find $\alpha= 1.1(1.5)$ and $\gamma=-0.4(2)$. The  errors 
reflect  large fluctuations in the values of the two parameters
$\alpha$ and $\gamma$ that do not allow us  to determine whether
the helix-coil transition  is a weak first  order or a strong
second order phase transition.
This problem was also observed in our earlier work on 
polyalanine \cite{AH99b,AFH01d} where we were also 
not able to establish clearly the order of the helix-coil transition.
However, our results illustrate the strength of this transition
that also leads to the pronounced peak in the specific heat observed 
in Fig.~3, and suggest a nucleation mechanism for $\alpha$-helix 
formation.  Our data are more decisive in the
case of the second transition, at $T=265$ K, which marks the
collapse and folding of the peptide. Here we find $\alpha=0.32(8)$
and $\gamma=0.36(2)$. These values indicate  a second-order transition
which is consistent with what one would expect for a transition between
extended and compact structures and imply that collapse and folding
of the Ala$_{10}$-Gly$_5$-Ala$_{10}$ is connected with long range 
correlations between the residues. 

Our above analysis of the thermodynamics of our peptide suggests that
Ala$_{10}$-Gly$_5$-Ala$_{10}$ folds in a 2 step process. The first step
is the formation of $\alpha$-helices and can be characterized by a  
helix-coil transition temperature $T_{hc} = 483\pm 8$ K. The formation
of $\alpha$-helices then restricts the possible configuration space. 
Energetically most favorable is the folding of two $\alpha$-helices  
(made out of the alanine residues)  into a hairpin.   This second 
step can be characterized by a lower folding temperature
 $T_f = 265\pm 7$ K. Note that this folding temperature is in the  
biological relevant temperature regime while helix-formation can 
also happen at much higher temperatures. The above described two step 
folding of our artificial peptide is reminiscent of the well known 
framework \cite{Ptitsyn,KB} and collision-diffusion model \cite{KW} 
of folding which also propose that local elements of native local 
secondary structure form independently of tertiary structure. These 
elements diffuse until they collide and coalesce to give a tertiary 
structure. In our case, the temperature region of $265 - 480$ K is the 
one where the thermal energy of the molecule does not allow 
coalescing of the helix-fragments that therefore form and decay. 
Some stabilization happens when these fragments try to form one 
extended helix, however, the inherent flexibility of the glycine 
residues, connecting the two alanine chains,  and the gain in 
Lennard-Jones energy lead instead at temperatures below $T_f$
to  a U-turn-like bundle of two (antiparallel) $\alpha$-helices 
connected by a turn of glycine residues as the most stable structure. 
Note that this picture is consistent with energy landscape theory 
and funnel concept \cite{Bryngelson87,Onuchic97}.  Fig.~5b depicts 
the appearance of a folding funnel at $T=270$ K towards our
``native structure'' `A'. The competing structure `B', that at
this temperature has a slightly higher  ($\approx 1 k_BT$) free 
energy ( see the 1 $k_BT$ contour line for structure `A' that is 
missing for conformer `B'), acts as a local trap. However, the free 
energy barriers of $\approx  6~k_BT$ can be  overcome at this
temperature by thermal fluctuations. Below that temperature the 
relative weight of structure `B' decreases (see Fig.7) and its
free energy difference to `A' increases: the energy landscape 
becomes even more funnel like (data not shown). The energy landscape 
of Fig.~5b allows for a multitude of folding pathways that all, however, 
 follow the above described two-step process.

An interesting question is how general the above obtained results  are.
A direct comparison with experimental data is difficult since
solvent effects were neglected in the simulation of 
 Ala$_{10}$-Gly$_5$-Ala$_{10}$ and most experiments study
solvated peptides. An exception are the techniques developed by 
Jarrold and collaborators for examination of gas-phase conformations 
of proteins and peptides \cite{jarrold}.  An experimental study
of  Ala$_{10}$-Gly$_5$-Ala$_{10}$ using these techniques is now planed.
The authors are not aware of experimental results for the solvated 
peptide. In order to compare our  work with experiments of 
other short helical peptides \cite{Brooks,Oas}, we therefore started 
now simulations of Ala$_{10}$-Gly$_5$-Ala$_{10}$ where the
solvation effects are approximated  by a solvent accessible surface
term \cite{inplan}. This will allows us also to test the dependency
of our results on the solvation model. Simulating another, slightly 
more complicated, artifical peptide, 
Ala$_{10}$-Gly$_5$-Ala$_{10}$-Gly$_5$-Ala$_{10}$  
that presumably will fold in a three-helix bundle, will  allow 
in addition a direct comparison with recent experimental work by 
Myers and Oas \cite{Oas}
where the relation between helix-formation and folding was studied 
for the 58-residue B domain of protein A. 

\section{conclusion}
In summary, we have performed multicanonical simulations with high
statistics of a simple artificial peptide, the 25 residue
 Ala$_{10}$-Gly$_5$-Ala$_{10}$. We found that this peptide 
folds into a specific  structure that is determined solely by
the intrinsic properties of the molecule (since solvent interactions
are absent in our simulations. In gas-phase, the peptide exhibits
two characteristic transitions. At $T_{hc}=483\pm 8$ K 
we observed a  helix-coil transition that is either a weak first 
order transition or a strong second order transition. Our results 
indicate that there is a second transition at $T_f=265\pm 7$ K, 
the folding transition, that is second order-like. These results 
suggest that folding of this peptide in gas-phase is a two-step 
process. In a first step, the alanine residues form independently 
helical segments which then afterwards in a second step assemble to 
a U-turn like structure of two antiparallel $\alpha$-helices
connected by a turn.  By using an implicit solvent model in
our simulations we started now to investigate whether the 
final structure or this two-step process  changes in the present of
water. 

 
\noindent
{\bf Acknowledgements}: \\
 U. Hansmann gratefully acknowledges support by  a research grant
from the National Science Foundation (CHE-9981874), and
N.A. Alves support by CNPq (Brazil).


 

\vfil
\newpage
\begin{table}[!h]
\renewcommand{\tablename}{Table}
\caption{Dihedral angles for the two configurations shown in Fig.~5}
\begin{center}
{\small
\begin{tabular}{|c|ccc|ccc|} 
State &\multicolumn{3}{c}{Conformer {\it A}} &
        \multicolumn{3}{c}{Conformer {\it B}}\\ \hline
$E_{tot}$ (Kcal/mol) & \multicolumn{3}{c}{-34.7} & \multicolumn{3}{c}{-23.9}\\
\hline
Residue & $\phi$ & $\psi$ & $\chi$ & $\phi$ & $\psi$ & $\chi$ \\
\hline
 Ala-1& -101.7 &  154.4 & -178.1 &  -55.2 &   -54.9 &    61.5\\
 Ala-2&  -65.4 &  -35.5 & -173.8 &  -70.6 &   -37.8 &    49.2\\
 Ala-3&  -68.2 &  -36.1 &  -61.7 &  -60.4 &   -30.8 &  -178.3\\
 Ala-4&  -70.6 &  -39.0 &  -69.7 &  -78.6 &   -37.6 &   -43.4\\
 Ala-5&  -64.3 &  -40.6 &  -54.1 &  -66.5 &   -28.8 &   176.5\\
 Ala-6&  -66.3 &  -38.3 &  175.1 &  -70.6 &   -50.4 &   -55.8\\
 Ala-7&  -69.5 &  -31.9 & -172.6 &  -64.2 &   -32.6 &  -178.6\\
 Ala-8&  -75.5 &  -31.6 &   57.2 &  -68.6 &   -43.4 &   -58.3\\
 Ala-9&  -62.6 &  -45.2 & -172.9 &  -67.8 &   -41.0 &  -164.8\\
 Ala-10&  -73.4 &  -54.4 &  -59.0 &  -60.8 &   -43.3 &    53.4\\
 Gly-11&  -91.9 &   65.0 &        &  -64.3 &   -40.5 &        \\
 Gly-12&  156.2 &  -80.6 &        &  -64.9 &   -44.9 &        \\
 Gly-13&  151.5 & -176.6 &        &  -64.6 &   -47.0 &        \\
 Gly-14&  -60.1 &  -35.5 &        &  -63.0 &   -33.0 &        \\
 Gly-15&  -63.5 &  -38.0 &        &  -72.6 &   -39.9 &        \\
 Ala-16&  -72.6 &  -34.6 &  -48.7 &  -70.3 &   -31.1 &    83.1\\
 Ala-17&  -69.4 &  -32.9 &  -52.2 &  -69.2 &   -39.2 &    63.3\\
 Ala-18&  -70.2 &  -38.4 &  -53.6 &  -68.4 &   -39.9 &    63.8\\
 Ala-19&  -70.5 &  -35.1 &  174.4 &  -66.1 &   -38.7 &   -58.5\\
 Ala-20&  -66.6 &  -40.4 &  179.5 &  -67.1 &   -41.9 &  -170.9\\
 Ala-21&  -68.7 &  -40.3 &  173.3 &  -66.0 &   -33.3 &    66.1\\
 Ala-22&  -61.2 &  -36.0 &   61.7 &  -73.9 &   -35.5 &    61.9\\
 Ala-23&  -71.1 &  -52.6 &  -51.3 &  -65.2 &   -37.8 &  -176.5\\
 Ala-24& -153.3 &  111.8 &   58.6 &  -73.0 &   -39.3 &   -39.5\\
 Ala-25&  -62.8 &  -62.6 & -173.5 &  -69.5 &    -8.8 &    71.0\\
\end{tabular}
}
\end{center}
\end{table}

\begin{table}[!h]
\renewcommand{\tablename}{Table}
\caption{\baselineskip=0.8cm Partition function zeros for
the two transitions observed for Ala$_{10}$-Gly$_5$-Ala$_{10}$.}
\begin{center}
\begin{tabular}{cccccccc}\\
\\[-0.3cm]
Re$(\beta_1)$ &$\tau_1$  &Re$(\beta_2)$ & $\tau_2$ &
Re$(\beta_3)$ &$\tau_3$  &Re$(\beta_4)$ & $\tau_4$        \\
\\[-0.35cm]
\hline
\\[-0.3cm]
 1.0463(46)   & 0.1307(53) & 1.0144(99) & 0.2112(55) &
 1.051(44)    & 0.310(36)  & 1.055(46)  & 0.347(41) \\
 1.855(21)    & 0.263(12)  & 1.991(25)  & 0.637(41) &
 2.057(53)    & 0.923(43)  & 2.070(43)  & 1.229(55) \\
\end{tabular}
\end{center}
\label{table 1}
\end{table}

\clearpage 
\newpage
{\huge Figure Captions:} \\
\begin{description}
\item[Fig.~1] The average helicity $<n_H>(T)$ of Ala$_{10}$-Gly$_5$-Ala$_{10}$ 
              as a function of temperature $T$. The corresponding values
              of the susceptibility $\chi (T)$ are ploted in the inset.
              Our data rely on a single
              multicanonical simulation of 8,000,000 Monte Carlo sweeps.
\item[Fig.~2] The average ECEPP/2 energy $<E_{tot}>$ and the corresponding
              partial energy terms, Coulomb energy $<E_C>$, Lennard-Jones
              term $<E_{LJ}>$, hydrogen-bond energy $<E_{hb}>$ and
              torsion energy $<E_{tor}>$, as a function of temperature
              $T$ for Ala$_{10}$-Gly$_5$-Ala$_{10}$.
\item[Fig.~3]  The specific heat $C(T)$ of  Ala$_{10}$-Gly$_5$-Ala$_{10}$ 
              as a function of temperature $T$.
\item[Fig.~4] The average end-to-end distance $<d_{e-e}>(T)$ as a function
              of temperature $T$ for  Ala$_{10}$-Gly$_5$-Ala$_{10}$.
\item[Fig.~5] Free energy landscape of Ala$_{10}$-Gly$_5$-Ala$_{10}$  as
              a function of helicity $n_H$ and end-to-end distance
              $d_{e-e}$ at temperature (a) $T_{hc}=480$ K  and 
              (b) $T_f =270$ K. The contour lines are drawn in
              multiples of $k_BT$ indicated in the key.
\item[Fig.~6] Lowest-energy conformation `A' of Ala$_{10}$-Gly$_5$-Ala$_{10}$
              (a) and the  conformation `B' (which has the largest helicity) 
              (b) as found in our multicanonical simulation of 8,000,000 
              Monte Carlo sweeps.
\item[Fig.~7] Relative weight  of the lowest-energy conformation `A' and
              conformation `B' (which has maximal helicity) as a function
              of temperature.
\end{description}



\begin{figure}[b]
\begin{center}
\begin{minipage}[t]{0.95\textwidth}
\centering
\includegraphics[angle=90,width=0.72\textwidth]{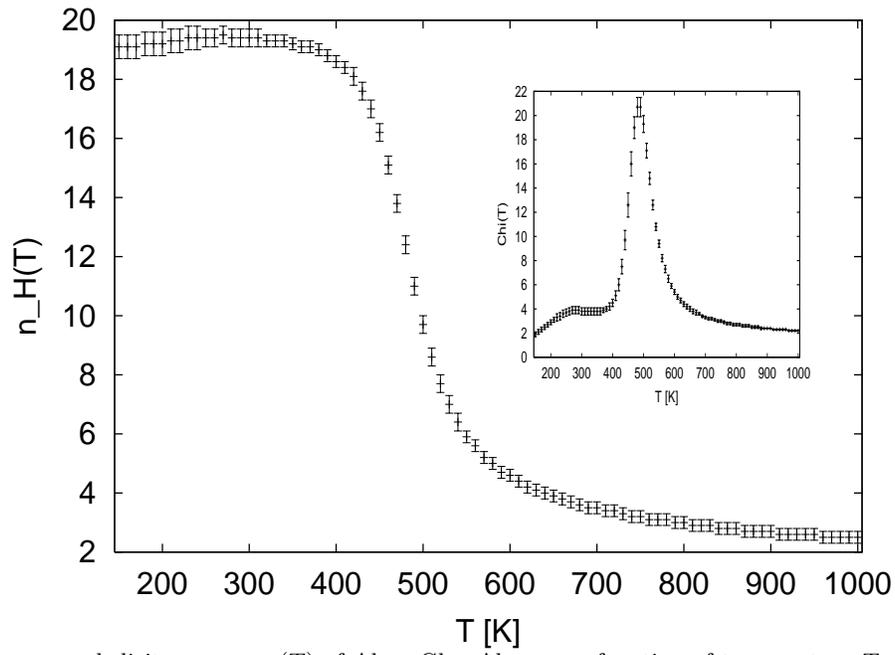}
\renewcommand{\figurename}{(Fig.1)}
\caption{The average helicity $<n_H>(T)$ of Ala$_{10}$-Gly$_5$-Ala$_{10}$
              as a function of temperature $T$. The corresponding values
              of the susceptibility $\chi (T)$ are ploted in the inset.
              Our data rely on a single
              multicanonical simulation of 8,000,000 Monte Carlo sweeps.}
\label{Fig. 1}
\end{minipage}
\end{center}
\end{figure}

\newpage
\cleardoublepage

\begin{figure}[b]
\begin{center}
\begin{minipage}[t]{0.95\textwidth}
\centering
\includegraphics[angle=-90,width=0.72\textwidth]{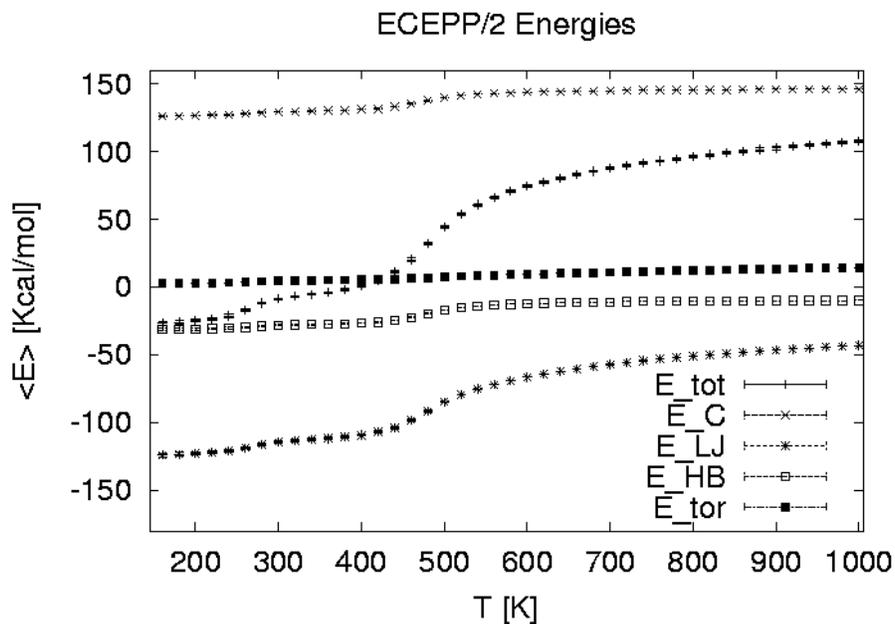}
\renewcommand{\figurename}{(Fig.2)}
\caption{ The average ECEPP/2 energy $<E_{tot}>$ and the corresponding
              partial energy terms, Coulomb energy $<E_C>$, Lennard-Jones
              term $<E_{LJ}>$, hydrogen-bond energy $<E_{HB}>$ and
              torsion energy $<E_{tor}>$, as a function of temperature
              $T$ for Ala$_{10}$-Gly$_5$-Ala$_{10}$. }
\label{Fig. 2}
\end{minipage}
\end{center}
\end{figure}

\begin{figure}[b]
\begin{center}
\begin{minipage}[t]{0.95\textwidth}
\centering
\includegraphics[angle=-90,width=0.72\textwidth]{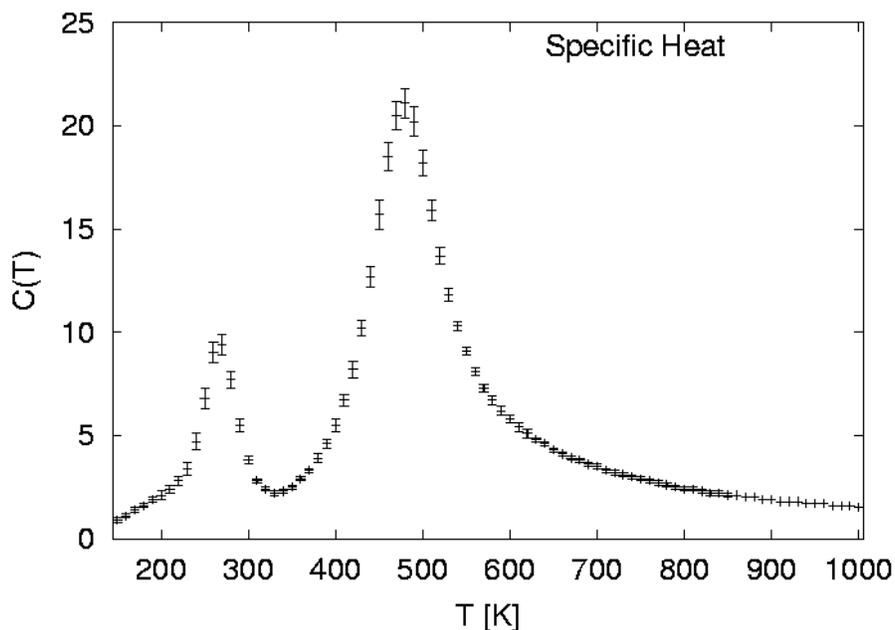}
\renewcommand{\figurename}{(Fig.3)}
\caption{ The specific heat $C(T)$ of  Ala$_{10}$-Gly$_5$-Ala$_{10}$ 
              as a function of temperature $T$. }
\label{Fig. 3}
\end{minipage}
\end{center}
\end{figure}

\newpage
\cleardoublepage

\begin{figure}[b]
\begin{center}
\begin{minipage}[t]{0.95\textwidth}
\centering
\includegraphics[angle=-90,width=0.72\textwidth]{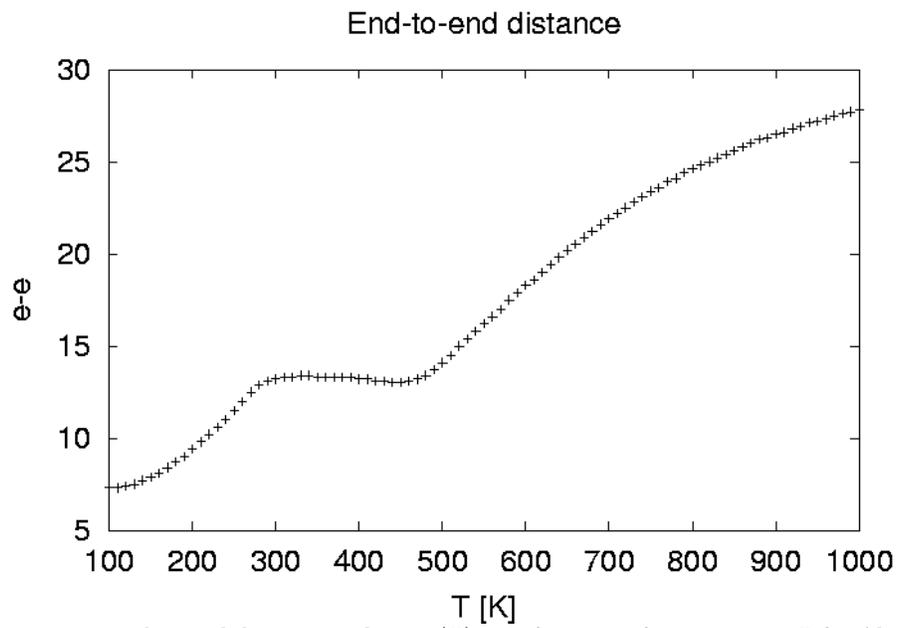}
\renewcommand{\figurename}{(Fig.4)}
\caption{The average end-to-end distance $<d_{e-e}>(T)$ as a function
              of temperature $T$ for  Ala$_{10}$-Gly$_5$-Ala$_{10}$.}
\label{Fig. 4}
\end{minipage}
\end{center}
\end{figure}

\newpage
\cleardoublepage

\begin{figure}[b]
\begin{center}
\begin{minipage}[t]{0.95\textwidth}
\centering
\includegraphics[angle=-90,width=0.72\textwidth]{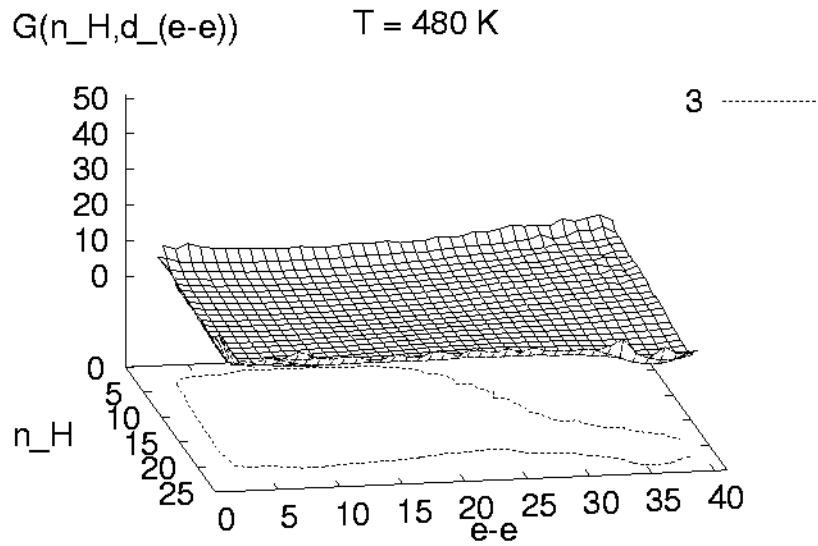}
\renewcommand{\figurename}{(Fig.5a)}
\caption{ Free energy landscape of Ala$_{10}$-Gly$_5$-Ala$_{10}$  as
              a function of helicity $n_H$ and end-to-end distance
              $d_{e-e}$ at temperature (a) $T_{hc}=480$ K  and 
              (b) $T_f =270$ K.}
\label{Fig. 5a}
\end{minipage}
\end{center}
\end{figure}

\begin{figure}[b]
\begin{center}
\begin{minipage}[t]{0.95\textwidth}
\centering
\includegraphics[angle=-90,width=0.72\textwidth]{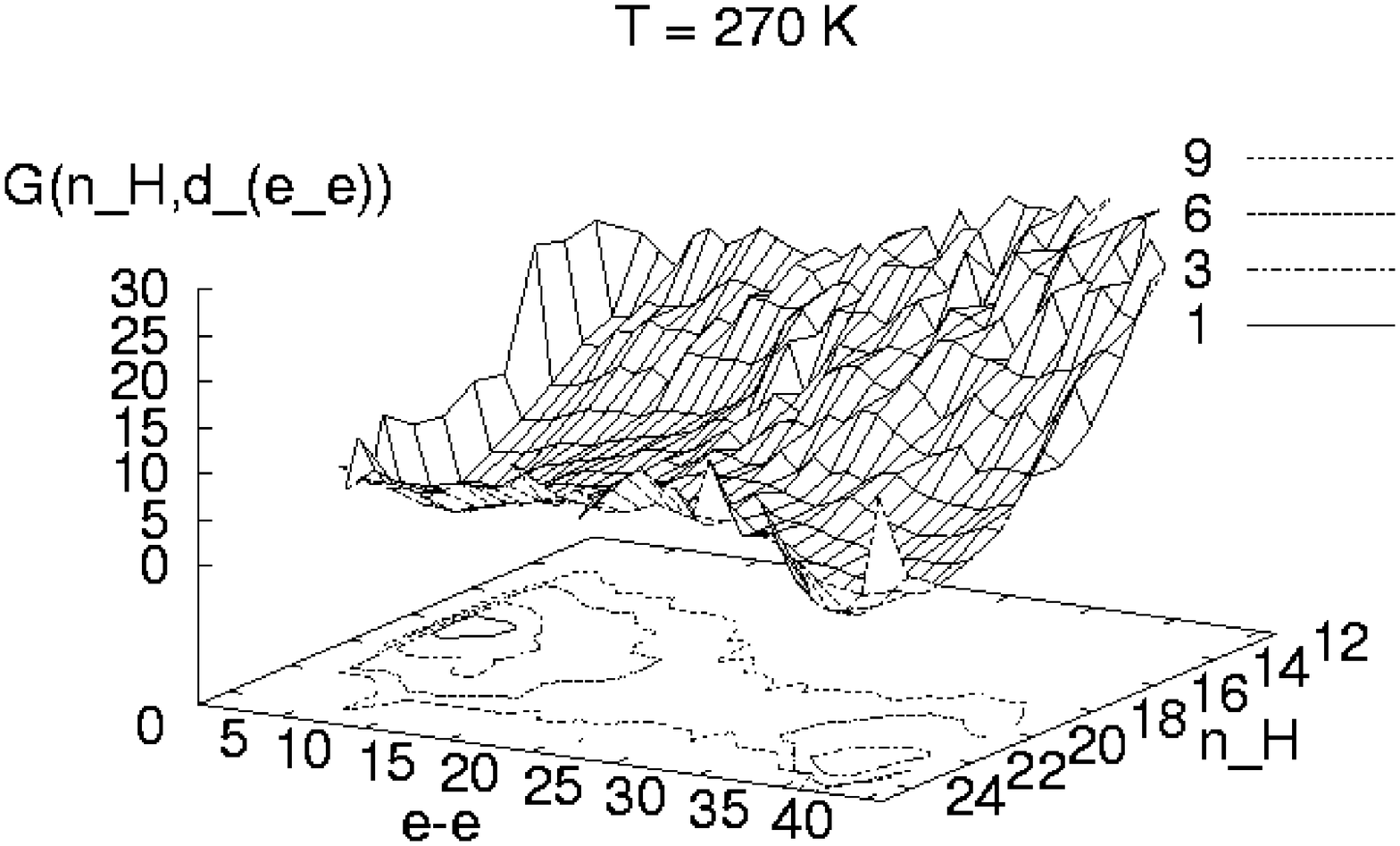}
\renewcommand{\figurename}{(Fig.5b)}
\caption{ }
\label{Fig. 5b}
\end{minipage}
\end{center}
\end{figure}

\newpage
\cleardoublepage

\begin{figure}[b]
\begin{center}
\begin{minipage}[t]{0.95\textwidth}
\centering
\includegraphics[angle=-90,width=0.72\textwidth]{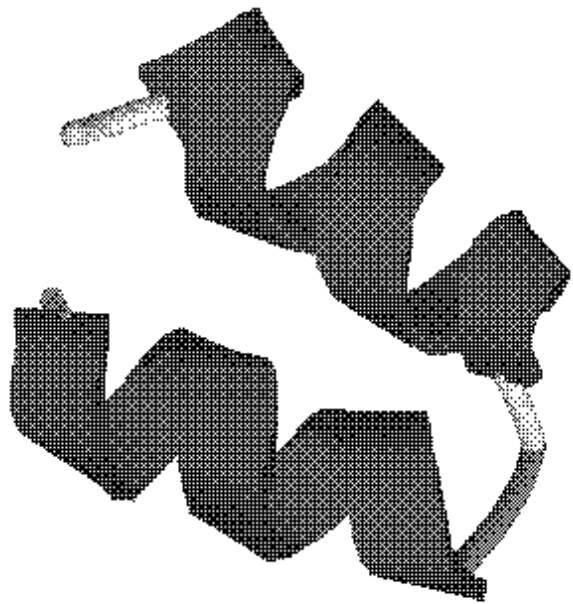}
\renewcommand{\figurename}{(Fig.6a)}
\caption{ Lowest-energy conformation `A' of Ala$_{10}$-Gly$_5$-Ala$_{10}$
              (a) and the  conformation `B' (which has the largest helicity) 
              (b) as found in our multicanonical simulation of 8,000,000 
              Monte Carlo sweeps.}
\label{Fig. 6a}
\end{minipage}
\end{center}
\end{figure}

\newpage
\cleardoublepage

\begin{figure}[b]
\begin{center}
\begin{minipage}[t]{0.95\textwidth}
\centering
\includegraphics[angle=-90,width=0.72\textwidth]{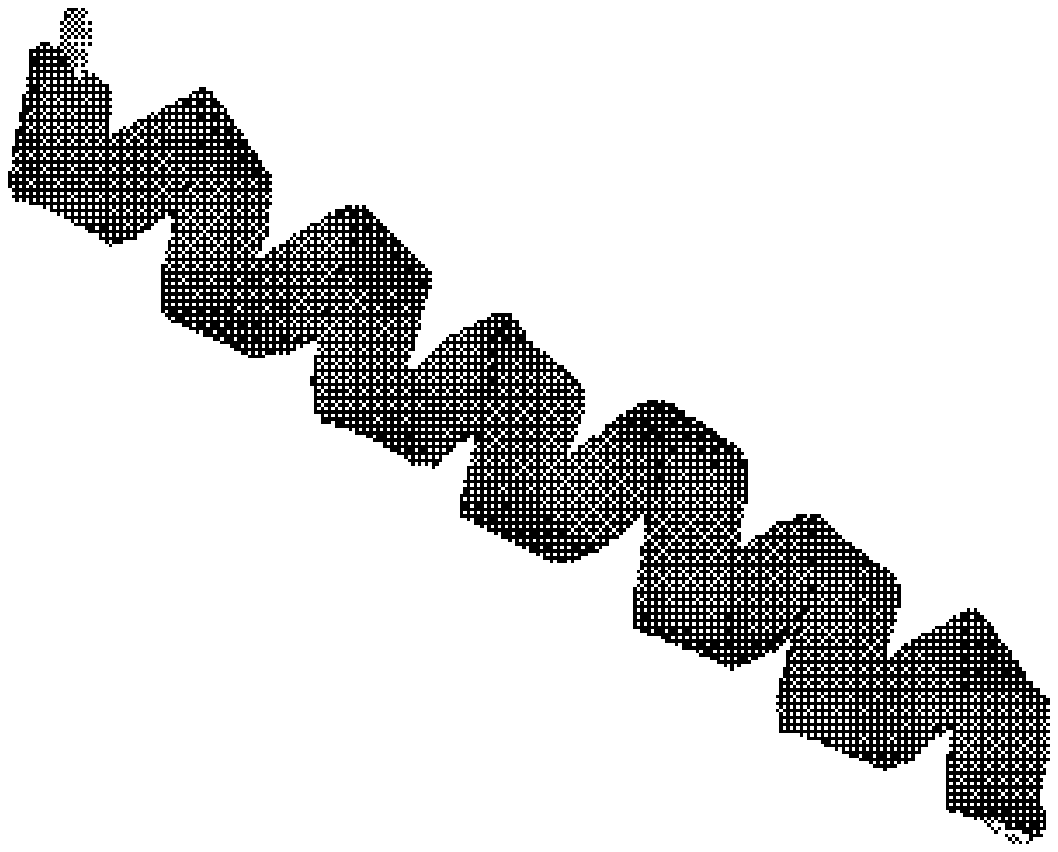}
\renewcommand{\figurename}{(Fig.6b)}
\caption{ }
\label{Fig. 6b}
\end{minipage}
\end{center}
\end{figure}

\newpage
\cleardoublepage

\begin{figure}[b]
\begin{center}
\begin{minipage}[t]{0.95\textwidth}
\centering
\includegraphics[angle=-90,width=0.72\textwidth]{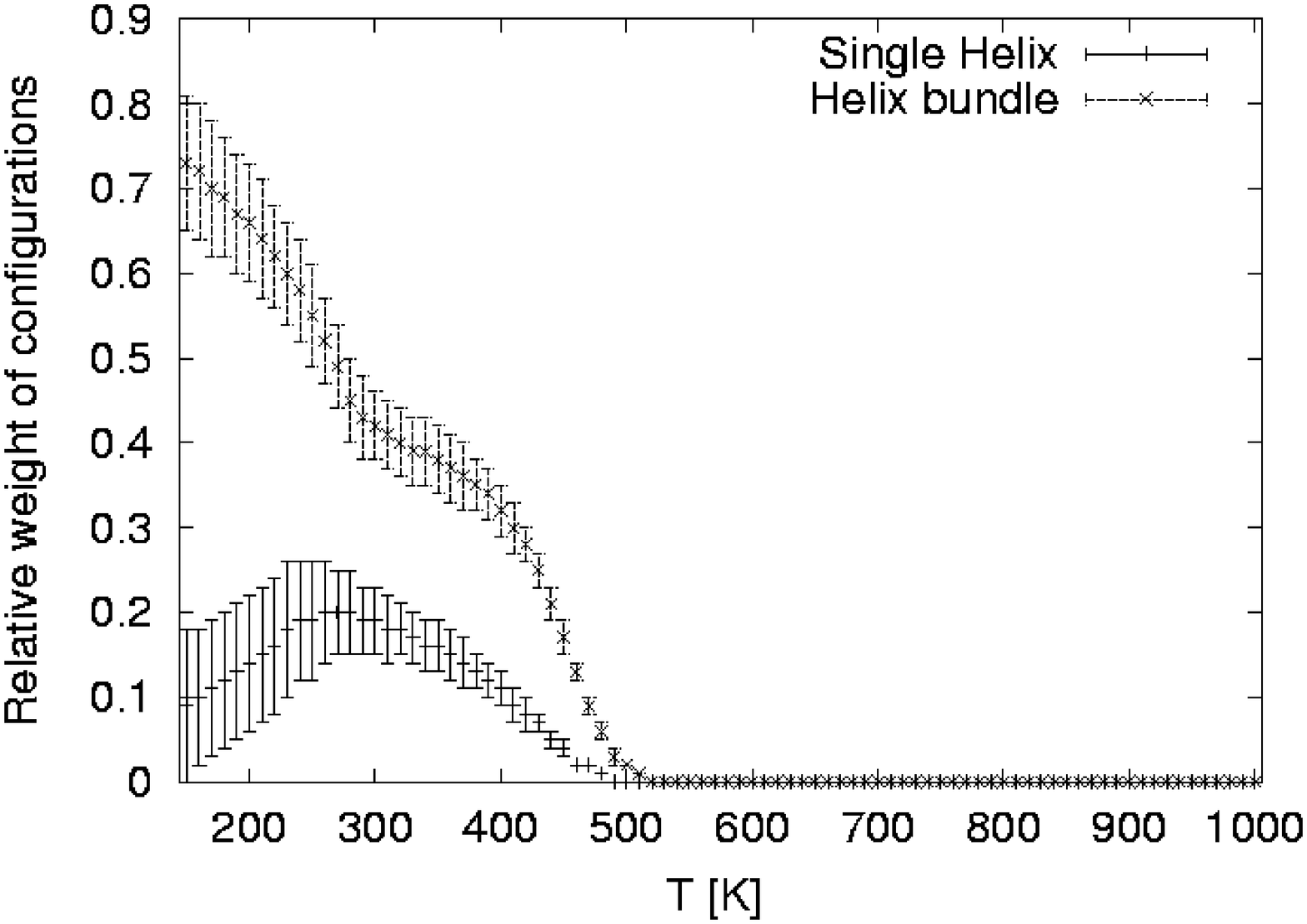}
\renewcommand{\figurename}{(Fig.7)}
\caption{ Relative weight  of the lowest-energy conformation `A' and
              conformation `B' (which has maximal helicity) as a function
              of temperature. }
\label{Fig. 7}
\end{minipage}
\end{center}
\end{figure}

\end{document}